\documentclass[12pt]{iopart}

\usepackage{graphicx}
\usepackage{epsfig}
\usepackage{color}

\newcommand{\REVIEW}[4]{#4 \textit{#1} \textbf{#2} #3}

\newcommand{\beq}{\begin{equation}}
\newcommand{\eeq}{\end{equation}}
\newcommand{\bea}{\begin{eqnarray}}
\newcommand{\eea}{\end{eqnarray}}
\newcommand{\bwt}{\begin{widetext}}
\newcommand{\ewt}{\end{widetext}}

\begin{document}

\title{From underdoped to overdoped cuprates: two quantum phase transitions}

\author{S G Ovchinnikov$^{1,2}$, E I Shneyder$^{1,3}$ and M M Korshunov$^{1,4}$}

\address{$^{1}$ L.V. Kirensky Institute of Physics, Siberian Branch of Russian Academy of Sciences, 660036 Krasnoyarsk, Russia}
\address{$^{2}$ Siberian Federal University, Krasnoyarsk, 660041, Russia}
\address{$^{3}$ Reshetnev Siberian State Aerospace University, Krasnoyarsk 660014, Russia}
\address{$^{4}$ Department of Physics, University of Florida, Gainesville, Florida 32611, USA}

\ead{korshunov@phys.ufl.edu}

\date{\today}

\begin{abstract}
Several experimental and theoretical studies indicate the existence of a
critical point separating the underdoped and overdoped regions of the
high-$T_c$ cuprates' phase diagram. There are at least two distinct proposals
on the critical concentration and its physical origin. First one is associated
with the pseudogap formation for $p < p^*$, with $p^* \approx 0.2$. Another one relies on the Hall effect measurements and suggests that the critical point and the quantum phase transition (QPT) take place at optimal doping, $p_{opt}
\approx 0.16$. Here we have performed a precise density of states calculation
and found that there are \textit{two} QPTs and the corresponding critical
concentrations associated with the change of the Fermi surface topology upon
doping.
\end{abstract}




\section{Introduction}
The mystery of high-$T_c$ superconductivity in layered cuprates is tightly
related to their common pattern of the doping-dependent transition from the
antiferromagnetic insulator at zero doping to the overdoped metal. A number of
experimental and theoretical studies indicate that the transition is not smooth and a critical point separates the underdoped (UD) and overdoped (OD) regions.
It is tempting to associate such critical point with the pseudogap formation
for $p < p^*$, with $p^*=0.19 \div 0.24$ \cite{rr1,r1,rr3,r2,r3}. No doubt that the proximity of the pseudogap and the superconductivity with two energy
scales, $T^*$ and $T_c$, is essential for high-$T_c$ superconductivity
\cite{r4}. On the other hand, the Hall effect measurements suggest that the
critical point and the quantum phase transition (QPT) take place at optimal
doping, $p_{opt} = 0.16$ \cite{rr7,r5}. To resolve this controversy, here we study the doping-dependent electronic structure of the single-layer cuprate like La$_{2-x}$Sr$_x$CuO$_4$ in the regime of strong electron correlation within the $t - t' - t'' - J^*$  model. By a very precise density of states (DOS) calculation we have found two QPT associated with the changes of the Fermi surface (FS) topology. At optimal doping, $x_{c1} = p_{opt} = 0.151$, the DOS reveals the logarithmic divergence while at the pseudogap QPT, $x_{c2} = p^* = 0.246$, there is a Heaviside-type step in the DOS.

Angle-resolved photo-emission spectroscopy (ARPES) reveals a change of the FS
topology from the small hole pockets to the large hole FS near the optimal
doping \cite{r6,r7}. This provides a link between the QPT and changes of the
FS. Here we apply the general Lifshitz ideas \cite{r8} on the QPT induced
by the FS transformations. But first of all, we will discuss how these
transformations are induced by doping.

It is easy to obtain the large FS in cuprates by a single-electron approach
like the local density approximation (LDA) or the tight-binding method.
However, to get the small hole pockets around the $(\pm \pi/2, \pm \pi/2)$
points of the Brillouin zone one have to go beyond the weak-coupling
approximations and take the strong electronic correlations into account. Such small pockets have been found in a doped antiferromagnetic (AFM) Mott insulator by exact diagonalization \cite{r9_1,r9_2} and Quantum Monte-Carlo calculations \cite{r10,rr13} for the finite clusters as well as by a
perturbative treatment of the infinite lattice \cite{r11,r12,rr16,r13} and
by using the slave-particles \cite{r9_3,r9_4}. According to these studies, after the long-range AFM order vanishes with increasing hole concentration $n_h = 1 + p$ (in La$_{2 - x}$Sr$_x$CuO$_4$, $p = x$), a short-range AFM order still persists even at optimal doping \cite{r14}. The short-range magnetic order
determines the self-energy and hole dispersion resulting in the small hole
pockets around the $(\pm \pi/2, \pm \pi/2)$ points in the UD cuprates, its
fluctuation results in the pseudogap formation \cite{r15,r16,r17}. Due to the
strong electronic correlations intrinsic for cuprates a theory of the electron
dynamics has to fulfill a ``no-double occupancy'' constraint. This constraint
is introduced explicitly in the mean-field theory of a $d$-type
superconductivity within the RVB approach \cite{rr22} for the $t - J$ model
\cite{r18}, and in the variational Monte-Carlo studies \cite{r19}.

\section{Method}
Contrary to the phenomenological approaches, like assuming that the second
order QPT exists at $p = p_c$ \cite{r20}, we deal with a microscopically
derived $t - t' - t'' - J^*$ model without free parameters. To properly fulfill the ``no-double occupancy'' constraint at every step of our calculations we use the Hubbard $X$-operators, $X^{hg} = \left| h \right\rangle \left\langle g
\right|$, where $\left| h \right\rangle$ and $\left| g \right\rangle$ are the
local eigenvectors corresponding to three states: one-hole states $\left|
\sigma \right\rangle$, $\sigma = \pm 1/2$, and the Zhang-Rice singlet, $\left|
S \right\rangle$, which is a two-hole state. The relation between $X$-operators and single-electron annihilation operators is given by $a_{f\sigma} = \sum\limits_{h,g} \gamma_{\sigma}(h,g) X_f^{hg}$, where the coefficients $\gamma_{\sigma}(h,g)$ determine the partial weight of the quasiparticle excitation $g \rightarrow h$ in the process of a particle annihilation on site $f$ with spin $\sigma$. The ``no-double occupancy'' constraint means the absence of direct excitations from and to the lower Hubbard band and the exclusion of the two-electron (zero-hole) state $\left| 0 \right\rangle = d^{10}p^6$ from the local Hilbert space. It is demonstrated straightforwardly by the exact calculation of the two-electron state occupation number that
$\left<n_{f\uparrow}n_{f\downarrow}\right>=\left<X_f^{00}\right>=0$; this
constraint is provided by the $X$-operators algebra. Nevertheless the virtual
interband (between the lower and upper Hubbard bands) hopping $t_{fg}^{12}$
results in the exchange interaction $J_{fg} = \left(t_{fg}^{12}\right)^2/U_{eff}$.

For La$_{2 - x}$Sr$_x$CuO$_4$, all intraband and interband hopping parameters
($t_{fg}^{11}$ and $t_{fg}^{12}$), single-site energies of holes in $p$- and
$d$-orbitals, and the charge transfer gap $U_{eff}$ have been calculated by the \textit{ab initio} LDA+GTB approach \cite{r21} which combines LDA and the
generalized tight-binding (GTB) method for strongly correlated systems. The low energy effective model is the $t - t' - t'' - J^*$ model where $J^*$ means that besides the Heisenberg exchange term a three-site correlated hopping $H_3$ is
also included, $H_{t - J^*} = H_{tJ} + H_3$, where
\begin{eqnarray}
\label{eq4}
 H_{tJ} &=& \sum\limits_{f, \sigma} (\varepsilon - \mu) X_f^{\sigma \sigma}
 + \sum\limits_f 2(\varepsilon - \mu) X_f^{SS} \nonumber\\
 &+& \sum\limits_{f \ne g, \sigma} \left[ t_{fg}^{11} X_f^{S \bar\sigma} X_g^{\bar\sigma S}
 + \frac{J_{fg}}{4} \left( X_f^{\sigma \bar\sigma} X_g^{\bar\sigma \sigma} - X_f^{\sigma \sigma} X_g^{\bar\sigma \bar\sigma} \right) \right], \nonumber\\
 H_3 &=& \sum\limits_{f \ne m \ne g, \sigma} \frac{t_{fm}^{12} t_{mg}^{12}}{U_{eff}}  \left( X_f^{\sigma S} X_m^{\bar\sigma \sigma} X_g^{S \bar\sigma} - X_f^{\sigma S} X_m^{\bar\sigma \bar\sigma} X_g^{S \sigma} \right). \nonumber
\end{eqnarray}
Here hole creation operator is now $\tilde a_{f\sigma}^\dag = 2\sigma X_f^{S
\bar\sigma}$ and its algebra is different from the bare fermion's one ($2\sigma = \pm 1$ for $\sigma=\uparrow,\downarrow$). The spin operators are also easily expressed via $X$-operators, $S_f^+= X_f^{\sigma
\bar\sigma}$, $S_f^z = \left( X_f^{\sigma \sigma} - X_f^{\bar\sigma \bar\sigma} \right)/2$.

Our approach is essentially a perturbation theory with the small parameter
$t/U$ contrary to the usual Fermi liquid perturbation expansion in terms of $U$ which is large in cuprates. We use a method of irreducible Green functions
which is similar to the Mori-type projection technique, with the zero-order
Green function given by the well-known Hubbard I approximation. Beyond it there are spin fluctuations. To provide a description of them, the self-energy was
calculated in the non-crossing approximation by neglecting vertex
renormalization that is equivalent to the self-consistent Born approximation
(SCBA) \cite{r23}. Resulting electron self-energy contains the space-time
dependent spin correlation function $C(\mathbf{q}, \omega)$ and results in the
finite quasiparticle lifetime, $\mathrm{Im}\Sigma(\mathbf{k},\omega) \ne 0$.
Note that at low temperatures $T \le 10$K the spin dynamics is much slower than the electron one. A typical spin fluctuation time, $10^{-9}$ sec, is much
larger than the electronic time $10^{-13}$ sec \cite{r26}; that is why we can
safely neglect the time dependence of the spin correlation function,
$C(\mathbf{q},\omega) \to C_\mathbf{q}$. The self-energy becomes static,
$\Sigma(\mathbf{k}, \omega) \to \Sigma(\mathbf{k})$, and we have
$\mathrm{Im}\Sigma = 0$. Note that $\Sigma(\mathbf{k}, \omega)$ here is the
object completely different from the one in the Fermi liquid approach because
the here it is build by the diagrams for the $X$-operators, not the
standard Fermionic annihilation-creation operators $a_{f\sigma}$. In the usual
Fermi liquid expansion dynamical self-energy definitely plays a crucial role in the lightly doped cuprates. Here, our theory starts from a different limit
where the lowest order approximation is represented by the Hubbard I solution.
The corrections to the strongly-correlated mean-field approach are small
because the starting point is already a reasonable approximation for the
Mott-Hubbard insulator. That is proved by the small effect of the frequency
dependence of the self-energy in Refs.~\cite{r22,r23}. Moreover, the doping-dependence of the FS is determined by $\mathrm{Re}\Sigma$, and it is
qualitatively similar in our approach \cite{r25} and in the approach which
properly takes $\mathrm{Im}\Sigma$ into account \cite{r22,r23}.

The vertex corrections to the self-energy are small far from the spin-density
wave or the charge-density wave instabilities, that is true for the moderate
doping. Our approximation for the self-energy is done in the framework of the
mode-coupling approximation which has been proved to be quite reliable even for systems with strong interaction \cite{r40,r41}. As shown in the spin-polaron
treatment of the $t-J$ model, the vertex corrections to the non-crossing
approximation are small and give only numerical renormalization of the model
parameters \cite{r42}.

Green function $\left\langle \left\langle \left. X_\mathbf{k}^{\bar\sigma S}
\right| X_\mathbf{k}^{S \bar\sigma} \right\rangle \right\rangle_\omega$ for a
hole moving on the background of short-range AFM order is
\begin{eqnarray}
\label{eq5}
 G(\mathbf{k},\omega) = \frac{(1+x)/2}{\omega - \varepsilon + \mu - \frac{1 + x}{2} t_\mathbf{k} - \frac{1 - x^2}{4} \frac{\tilde{t}_\mathbf{k}^2}{U_{eff}} + \Sigma(\mathbf{k})},
\end{eqnarray}
where
\begin{eqnarray}
\label{eq6}
 \Sigma(\mathbf{k}) = -\frac{2}{1 + x} \frac{1}{N} \sum\limits_\mathbf{q} \left\{ \left[ t_{\mathbf{k} - \mathbf{q}} - \frac{1 - x}{2} J_\mathbf{q}  + \frac{1 - x}{2} \frac{\tilde{t}_{\mathbf{k} - \mathbf{q}}^2}{U_{eff}} \right.\right. \nonumber\\
 - \left.\left. \frac{1 + x}{2} \frac{2 \tilde{t}_\mathbf{k} \tilde{t}_{\mathbf{k} - \mathbf{q}}}{U_{eff}} \right] \left( \frac{3}{2}C_\mathbf{q} + K_{\mathbf{k} - \mathbf{q}} \right) - \frac{1 + x}{2} \frac{\tilde{t}_\mathbf{q}^2}{U_{eff}} K_\mathbf{q} \right\}.
\nonumber
\end{eqnarray}
Here, $t_\mathbf{k}$ and $\tilde{t}_\mathbf{k}$ are the Fourier transforms of
hoppings $t_{fg}^{11}$ and $t_{fg}^{12}$, respectively. The self-energy is determined by static spin correlation function $C_{0n} = \left\langle {S_0^+ S_n^-} \right\rangle$ and kinetic correlation function $K_{0n} = \sum\limits_\sigma \left\langle \tilde a_{0\sigma }^\dag \tilde a_{n\sigma} \right\rangle$ between sites 0 and $n$. These correlation functions and their Fourier transforms $C_\mathbf{q}$ and $K_\mathbf{q}$ represent the AFM short-range order and the valence-bond order, respectively. In contrast to
approach of Ref.~\cite{r23}, we calculate these correlation functions
self-consistently up to $n = 9$  (ninth coordination sphere) together with the
chemical potential $\mu$. To get the spin correlation function we also obtain
the spin Green function $\left\langle \left\langle \left. X_\mathbf{q}^{\sigma
\bar\sigma} \right| X_\mathbf{q}^{\bar\sigma \sigma} \right\rangle
\right\rangle_\omega$ in a spherically-symmetric spin liquid
state~\cite{rr30,r24} with $\left\langle S^z \right\rangle = 0$ and the equal
correlation functions for each spin component, $\left\langle S_0^+ S_n^-
\right\rangle = 2 \left\langle S_0^z S_n^z \right\rangle = C_{0n}$. Both
$C_{0n}$ and $K_{0n}$ are essentially doping-dependent and $C_{0n}$ decrease
with the doping~\cite{r25}. While the nearest neighbor function $C_{01}$ is finite for all studied $x$ up to $x = 0.4$ with a kink at $x = p^* = 0.24$, more distant spin correlations fall down to zero for $x > p^*$.

\begin{figure}
\begin{center}
 \includegraphics[angle=0,width=0.7\columnwidth,clip=true]{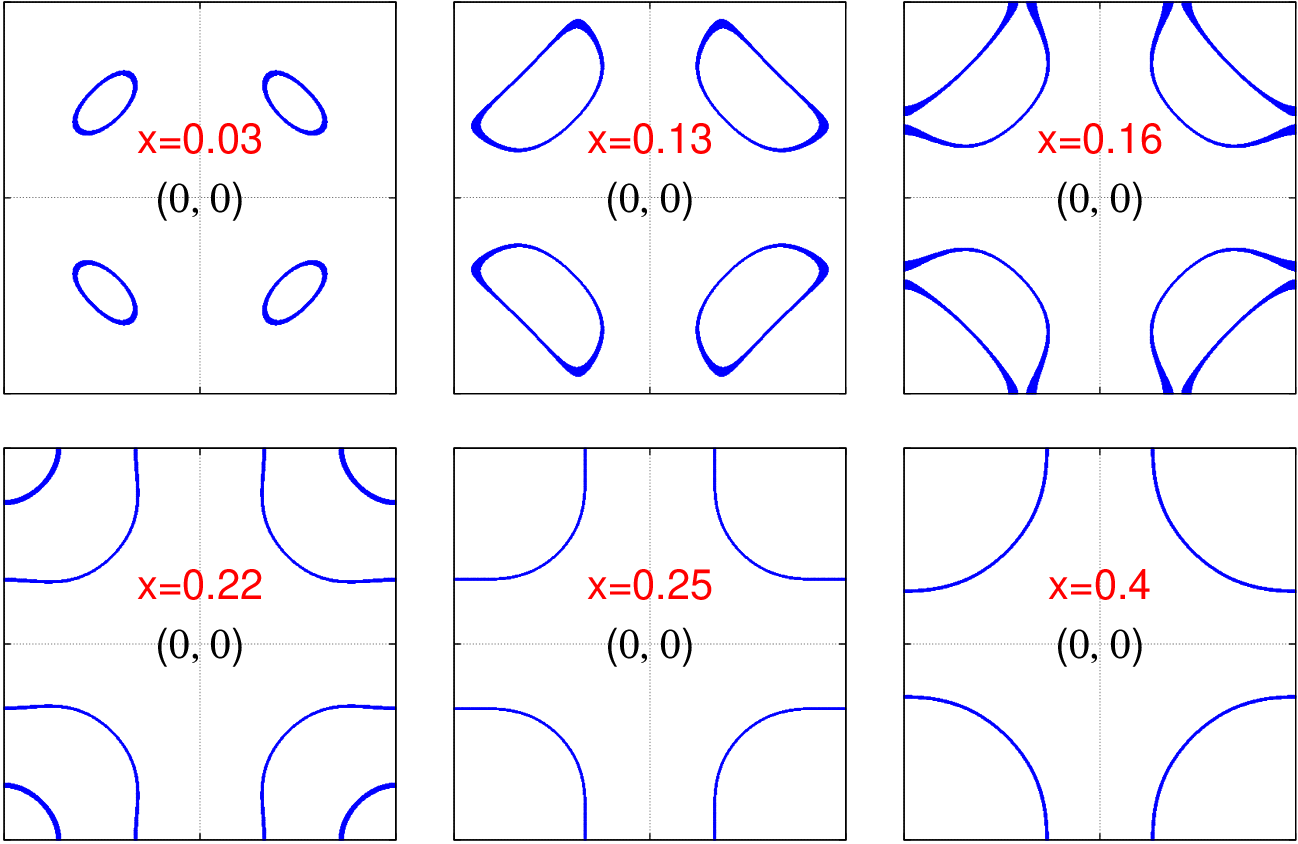}
 \caption{(Color online) Mean-field Fermi surface transitions with doping $x$ as calculated from poles of Eq.~\ref{eq5}. There are two topological changes: first one between $x=0.13$ and $0.16$, and second one between $0.22$ and $0.25$; see Ref.~\cite{r25} for detailed discussion.}
\label{figFS}
\end{center}
\end{figure}
The calculated FS twice changes its topology with doping \cite{r25}, see
Fig.~\ref{figFS}. Small hole pockets around $(\pm \pi/2, \pm \pi/2)$ points are present at small doping; then they increase in size and touch each other in the non-symmetric points $\mathbf{k} = \pm \pi (1, \pm 0.4)$ at $x_{c1} = p_{opt} = 0.151$. Above $p_{opt}$, there are two FSs around $(\pi, \pi)$ with outer being a hole-like and inner being an electron-like. The electron FS collapsed at
$x_{c2} = p^* = 0.246$, and at $x > p^*$ we have only one large hole surface
around $(\pi, \pi)$. Similar conclusion on the coexistence of hole and electron FS at some intermediate doping have been also drawn recently \cite{r27,r28},
and earlier for the spin-density wave sate of the Hubbard
model~\cite{rSachdev}.

It should be stressed that the standard DOS calculations with routine precision ($400 \times 400$ points in the quarter of the Brillouin zone) which we used before to solve the $T_c$ equation for magnetic mechanism of $d_{x^2 - y^2}$-wave pairing~\cite{r30} is not enough to find the effect of QPT on DOS. To get the results presented below we used $10^4 \times 10^4$ $k$-points which lead to the increase of precision by 625 times.

\section{Results}
From the previous consideration it follows that the FS topological transitions in cuprates are induced by doping and they are due to the non-rigid band behavior of the quasiparticles in the strongly correlated systems. According to the general Lifshitz analysis \cite{r8} for the three dimensional (3D) system, a change of topology at the energy $\varepsilon = \varepsilon_c$ either by appearance of a new segment (like we found at $p^*$) or by change of its connectivity (like at $p_{opt}$) would result in the additional DOS, $\delta N(\varepsilon) \sim \left( \varepsilon - \varepsilon_c \right)^{1/2}$, and the change in the thermodynamic potential, $\delta\Omega \sim \left(\varepsilon_F - \varepsilon_c \right)^{5/2}$ (the QPT of the 2.5-order), where $\varepsilon_F$ is the Fermi energy. However, due to the strong anisotropy of electronic and magnetic properties, cuprates are quasi-2D and not isotropic 3D systems. The electron hopping perpendicular to the CuO$_2$ layers in a single-layer La$_{2 - x}$Sr$_x$CuO$_4$ (LSCO), Bi$_2$Sr$_2$CuO$_{6 + \delta}$ (Bi2201), etc. is negligibly small. We do not consider here YBa$_2$Cu$_3$O$_{7 - \delta}$ (YBCO) and Bi$_2$Sr$_2$CaCu$_2$O$_{8 + \delta}$ (Bi2212) with two CuO$_2$ layers in the unit cell where the bilayer splitting of the FS appears, so we calculate DOS for the electrons in the doped single CuO$_2$ layer.

\begin{figure}
\begin{center}
 \includegraphics[angle=0,width=0.7\columnwidth]{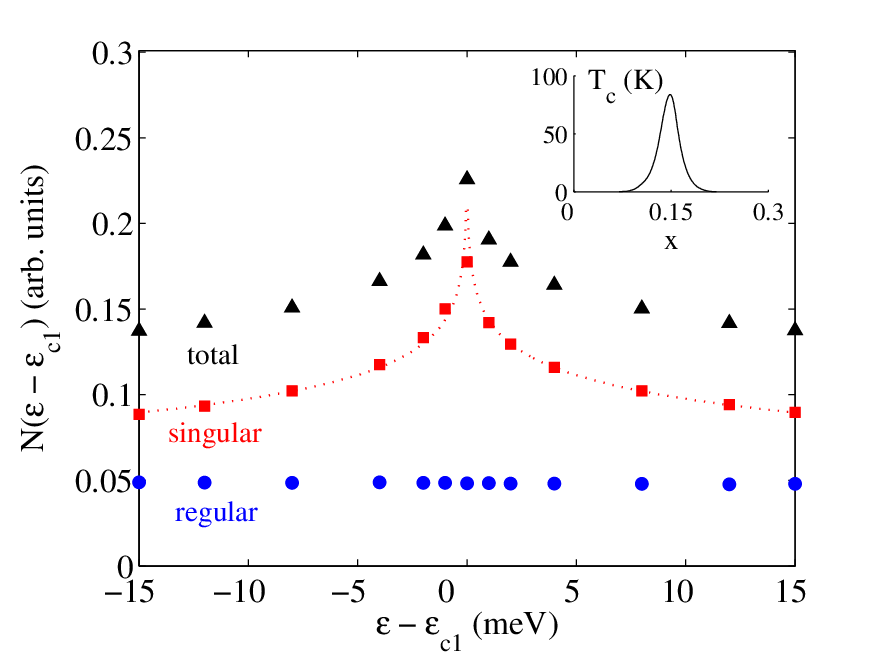}
 \caption{(color online) Regular, singular, and total density of states $N\left(\varepsilon-\varepsilon_{c1}\right)$ near the optimal doping, $\varepsilon_{c1}=\varepsilon_F(p_{opt})$, as calculated from the Green function (\ref{eq5}). Dotted line shows the logarithmic fitting. In the inset, the doping dependence of the superconducting critical temperature $T_c(x)$ is shown; the optimal doping is $0.151$. Note that the energy $\varepsilon - \varepsilon_{c1}$ is the energy of holes.}
\label{fig1}
\end{center}
\end{figure}
\begin{figure}
\begin{center}
 \includegraphics[angle=0,width=0.7\columnwidth]{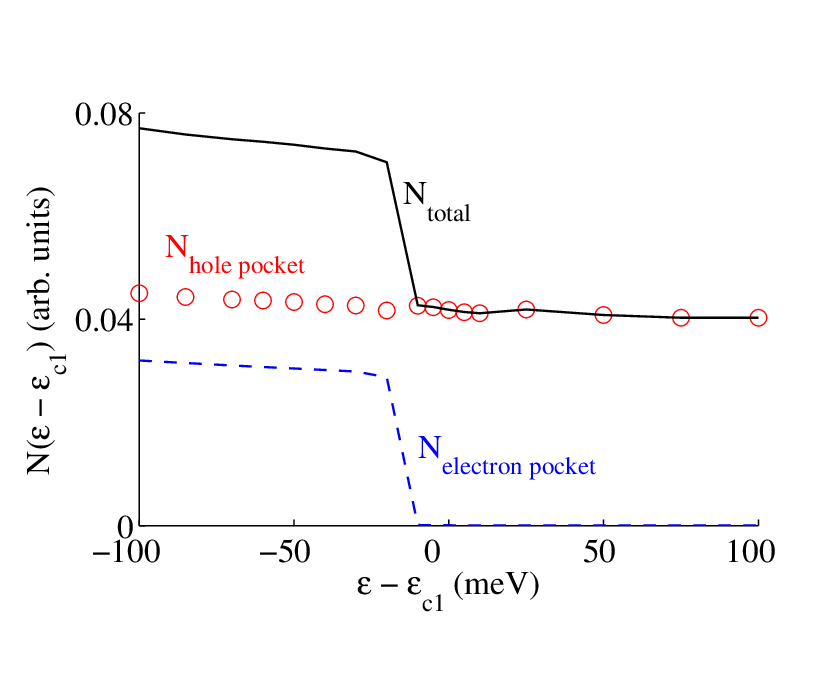}
 \caption{(color online) Regular (hole pocket), singular (electron pocket), and total density of states $N\left(\varepsilon - \varepsilon_{c1}\right)$ near the pseudogap critical point $\varepsilon_{c2} = \varepsilon_F(p^*)$. Below $p^* = 0.24$ ($\varepsilon < \varepsilon_{c2}$) a singular step-like contribution to the total DOS appears. Note that the energy $\varepsilon - \varepsilon_{c1}$ is the energy of holes.}
\label{fig2}
\end{center}
\end{figure}
The change of the FS topology at $x_{c1} = p_{opt}$ results in the logarithmic
divergence of DOS (Fig.~\ref{fig1}), while the emergence of the new
electron-like pocket below $x_{c2} = p^*$ results in a step in DOS
(Fig.~~\ref{fig2}). The total DOS is a sum of the singular and regular
contributions. We would like to stress that both logarithmic and step DOS singularities are in perfect agreement with the general properties of the van Hole singularities for the 2D electrons \cite{rr35}. Contrary to the 3D systems, the thermodynamical potential for the 2D electrons has a singular contribution $\delta\Omega \sim \left( \varepsilon_F - \varepsilon_c \right)^2$ for the step singularity and $\delta\Omega \sim \left( \varepsilon_F - \varepsilon_c \right)^2 \ln\left| \varepsilon_F - \varepsilon_c \right|$ for the logarithmic singularity \cite{r29}. Thus QPT at $x_{c2} = p^*$ is of the second order, while at $x_{c1} = p_{opt}$ the singularity is stronger. It is immediately follows that the Sommerfeld parameter $\gamma$ in the electronic heat capacity $\gamma = C_e/T$ has also a singular step contribution at $x \le p^*$, and
\begin{equation}
 \delta\gamma \propto \ln\left( \varepsilon_F - \varepsilon_c \right) \propto \ln\left| x - x_{opt} \right|
\label{eqsing}
\end{equation}
near $x_{c1} = p_{opt}$. Similar divergence in the specific heat was found
within the dynamical cluster approximation for the Hubbard model \cite{Mikelsons2009}.

\begin{figure}
\begin{center}
 \includegraphics[angle=0,width=0.7\columnwidth]{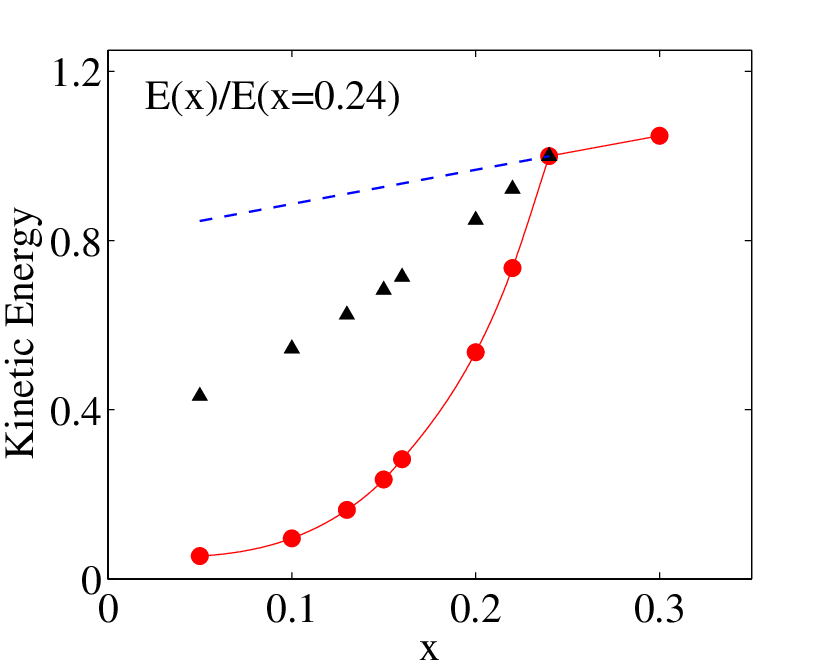}
 \caption{(color online) The doping dependence of the dimensionless kinetic energy $E_{kin}(p)/E_{kin}(p^*)$. Calculated dependence shown by the filled (red) circles. Above $p^*$ it obeys a conventional law and is proportional to $(1 + p)$. The extrapolation of this law to the region $p < p^*$ (blue dashed line) emphasizes the depletion of part of the kinetic energy in the pseudogap region. Calculation for the idealized triangular pseudogap model (\ref{eq2}) is shown by the filled triangles.}
\label{fig3}
\end{center}
\end{figure}
To check whether the coincidence of $x_{c1}$ with $p_{opt}$ and $x_{c2}$ with
$p^*$ is occasional or not, we have calculated the superconducting critical
temperature dependence $T_c(x)$ in the same model \cite{r30} and the kinetic
energy as a function of doping. The $T_c(x)$ dependence is an inverse parabola
with the maximum at $x_{opt}$ (see inset in Fig.~\ref{fig1}), which indeed
equals to $x_{c1}$. Note that it is not a coincidence since like in the BCS
theory the maximum in $T_c(x)$ is determined by the maximum DOS, and at
$x_{c1}$ we have a logarithmic singularity. Kinetic energy, $E_{kin} = \sum\limits_n t_{0n}^{11} K_{0n}$, reveals a remarkable kink at $x_{c2} = p^*$
(Fig.~\ref{fig3}). Above $p^*$, $E_{kin}(p)/E_{kin}(p^*) \sim 1 + p$ that is
expected for a conventional 2D metal with the hole concentration $n_h = 1 + p$ and $E_{kin} \sim \varepsilon_F \sim n_h$. The extrapolation of this law below
$p^*$ (shown in Fig.~\ref{fig3} by the blue dashed line) reveals that actual
$E_{kin}$ is smaller. We associate this depletion of the kinetic energy with
the pseudogap formation and try to fit it with a simple free electron gas with
a triangular pseudogap DOS (Loram-Cooper model \cite{rr1,r1}):
\begin{equation}
\label{eq2}
 N(\varepsilon) = \left\{ \begin{array}{l}
              g, \left| \varepsilon - \varepsilon_F \right| > E_g \\
              g \frac{\left| \varepsilon - \varepsilon_F \right|}{E_g},
               \left| \varepsilon - \varepsilon_F \right| < E_g.
                         \end{array} \right.
\end{equation}
Here $E_g = J (p^* - p)/p^*$ is a doping dependent pseudogap and $J$ is the
nearest-neighbor exchange parameter. This fitting is shown by the filled
triangles and it reflects some decrease of the kinetic energy due to the
pseudogap but do not provide a quantitative agreement. Apparently, better
fitting is given by the exponential law, $E_{kin}(p)/E_{kin}(p^*) = \exp \left[ -4 E_g(p)/J \right]$. This analysis confirms that the QPT at $x_{c2}$ is indeed related to the pseudogap and the coincidence of $x_{c2}$ and $p^*$ is not occasional.

\begin{figure}
\begin{center}
 \includegraphics[angle=0,width=0.7\columnwidth]{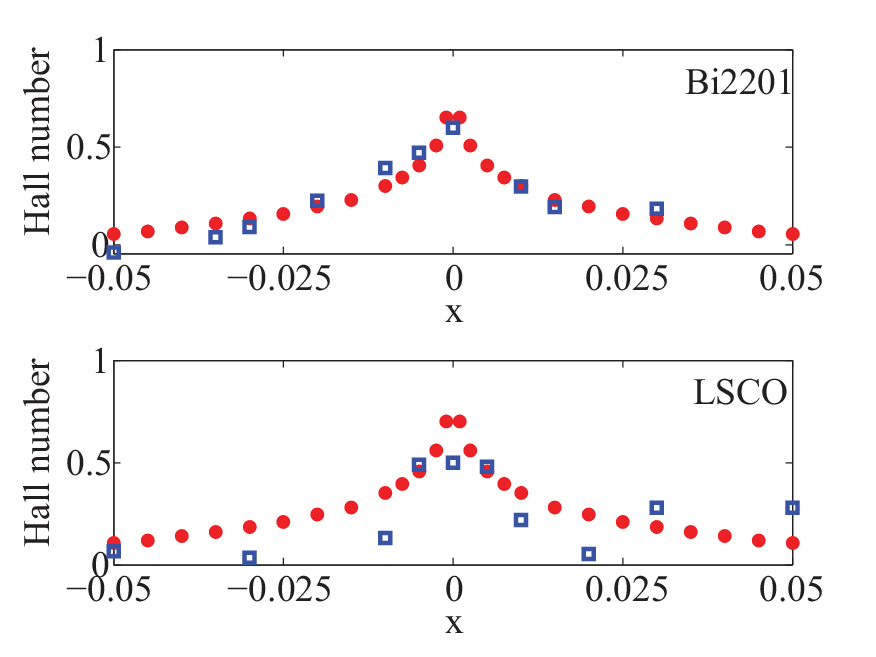}
 \caption{(color online) Comparison of experimental (blue squares) singular Hall coefficient \cite{r5} for Bi$_2$Sr$_{0.51}$La$_{0.49}$CuO$_{6 + \delta}$ (a) and for La$_{2 - x}$Sr$_x$CuO$_4$ (b) and our calculated (red filled circles) singular DOS, $N_{sing}\left(\varepsilon_F(x)\right)$, near the optimal doping as follows from Eq.~(\ref{eqsing}). Agreement with results on bulk single crystals (a) is better than with results on thin films (b).}
\label{fig4}
\end{center}
\end{figure}
A singular contribution to the Hall coefficient near the optimal doping has
been measured for Bi$_2$Sr$_{0.51}$La$_{0.49}$CuO$_{6 + \delta}$ single
crystals and for La$_{2 - x}$Sr$_x$CuO$_4$ thin films under strong magnetic
field of 60T \cite{r5}. According to our theory, these extra carriers are
induced by the singular DOS. To continue discussion in terms of the critical
points, not the critical energies, we note that near the critical point
$\varepsilon_F(x) - \varepsilon_{c1} = k (x - x_{opt})$. In Fig.~\ref{fig4} we
plot the calculated singular DOS $N_{sing}(z)$, $z = x - x_{opt}$, together
with the singular contribution to the Hall coefficient,
$n_{Hall}(1.5\mathrm{K}) - n_{Hall}(100\mathrm{K})$ \cite{r5}. The optimal
doping in the LSCO thin film $x_{opt} = 0.17$ is shifted from the bulk value
$x_{opt} = 0.15$ in the Bi2201 which may be due to the strains in the films.
The general agreement of the calculated singular DOS and Hall data provides
further support for our analysis.

\section{Discussion}
Now we are going to compare our results with the other relevant studies and
discuss the retardation effects for the electronic self-energy. These effects
determine $\mathrm{Im}\Sigma(\mathbf{k},\omega)$ and hence the quasiparticle
spectral weight and line width. Our approach allows to go beyond the static
limit and to get the frequency dependent real and imaginary parts of the
self-energy by the Mori-type projection technique; for the Hubbard model
calculations of such type have been done in Ref.~\cite{r23}. The authors of
Ref.~\cite{r23} get very similar concentration dependence of the Fermi surface
(see Fig.~6 in their paper) as we have. This agreement proves that
$\mathrm{Im}\Sigma(\mathbf{k},\omega)$ is not so important for the shape of the Fermi surface. Nevertheless it is important for the spectral weight. In
particular, the spectral weight of the inner pocket around $(\pi,\pi)$ is small due to the finite quasiparticle lifetime. Thus the ARPES intensity for this
pocket is small and that may be the reason why it has never been observed by
ARPES.

The energy dependence of the electron self-energy is crucial and determines the Mott-Hubbard transition in the Hubbard model as was convincingly demonstrated
by the dynamical mean-field theory (DMFT) \cite{r43}. Cluster generalization of DMFT \cite{r44,r45,r46,r47} is necessary to study electron correlations in a
two-dimensional CuO$_2$ layer where the nearest neighbor spin correlations
require the momentum dependent self-energy. The cellular DMFT (CDMFT) method
provides $k$-dependent self-energy and results in the phase
diagrams that have features similar to the ones experimentally observed in
cuprates \cite{r48,r49,r50,r51,r52}. Recently, the exact diagonalization
version of CDMFT (CDMFT+ED) was used to study the electronic structure of the doped Mott-Hubbard insulator \cite{r53,r54}. The sequence of the FS transformations with doping in Refs.~\cite{r53,r54} is very similar
to ours. At a small doping $x$, four hole pocket expand with $x$ until they touche the Brillouin zone boundary ($|k_x|$ or $|k_y|$ equal to $\pi$). Then at $x_{c1}=p_{opt}$ they merge into two concentric Fermi surfaces around
$(\pi,\pi)$. With further doping, the smaller surface disappears leaving a
large hole-like FS which lately transforms into a normal electron-like one
through one more Lifshitz QPT. In spite of many differences in details (for
example both poles and zeros of the Green function are obtained in
Refs.~\cite{r53,r54}) the similarity of the FS transformations in our work and
papers \cite{r53,r54} proves the validity of our approach at least at low
temperatures. We believe that it is the simplest approach that allows to obtain the Fermi surface transformation from the lightly doped Mott insulator up to
the Fermi liquid. Nevertheless our static approximation cannot treat the
quasiparticle spectral weight (see discussion of Ref.~\cite{r23} above). It
does not work in the Fermi liquid regime either. We start our perturbation
theory with the small self-energy in the atomic limit and then the self-energy
will be large in the band limit.

One more agreement between our work and the dynamical cluster approximation is
the $T^2 \log T$ singularity in the thermodynamic potential at $x_{c1}$ in
Ref.~\cite{Mikelsons2009} and our $z^2 \log z$ [see Eq.~(\ref{eqsing})] at the
Lifshitz QPT. At zero temperature, $z$ is given by the energy difference of the Fermi level and the critical energy that is proportional to $(x-x_{c1})$. At
finite temperature $z \propto T$.

\section{Conclusion}
We have shown that there are \textit{two} critical points in the cuprate's
doping dependence. The first one is related to the change of the FS
connectivity and logarithmic divergences of DOS and of electronic heat capacity parameter $\gamma$ at the optimal doping $p_{opt} = 0.151$. Also, we associate
this QPT with the experimentally observed singular doping dependence of the
Hall coefficient \cite{r5}. Moreover, the logarithmic enhancement of DOS leads
to the maximum in the doping dependence of superconducting critical temperature $T_c$ at the same critical point $x = p_{opt}$. This is in agreement with the
previous calculations of $T_c$ for the magnetic mechanism of $d_{x^2 - y^2}$
pairing. The second QPT is associated with the collapse of the electron-like FS pocket at $p \to p^* = 0.246$ and results in the step singularities in DOS and
in Sommerfeld parameter $\gamma$. We have found the depletion of the hole's
kinetic energy below $p^*$ and ascribe it to the pseudogap formation at $p <
p^*$. Thus the two energy scales in cuprates measured by $T_c$ and $T^*$ are
both related to the QPTs and to the changes of the cuprate's electron structure with doping. The very existence of both logarithmic and step singularities in DOS are in perfect agreement with the general properties of the van Hole singularities for the 2D electron systems. But the concentrations of doping at which these singularities approach the Fermi level and start to govern the behavior of the system are determined by the strong electronic correlations and scattering on the associated short-range AFM order.

Note that our analysis is appropriate for cuprates that have one CuO$_2$ layer
in the unit cell. The question arise whether the model parameters and corresponding critical concentrations are the same for e.g. LSCO and Bi2201? In the conventional single electron tight-binding model with the rigid band the
hopping parameters depend on doping significantly. That is why the ratio $t'/t$ extracted from ARPES is usually different for Bi2201 and LSCO. In general, the
hopping parameters depend on the interatomic distance that is almost the same
in these two crystals. That is why we use the same parameters for all doping
concentrations. The doping dependence of the band structure and its non-rigid
behavior comes up as the effect of strong electronic correlations.

\ack
We would like to thank S. Sakai for useful discussions. The authors acknowledge support by the Russian Foundation for Basic Research (grant N 09-02-00127), by
the Integration Program of SBRAS N40, the Presidium RAS Program 5.7, President
of Russia (grant MK-1683.2010.2), FCP Scientific and Research-and-Educational
Personnel of Innovative Russia for 2009-2013 (GK P891), and in part by the
National Science Foundation under Grant NSF PHY05-51164.

\end{document}